\documentstyle[psfig]{mn2e}

\title[RR Lyrae-based calibration of the Globular Cluster Luminosity Function]{RR Lyrae-based calibration of the Globular Cluster Luminosity Function}

\author[M. Di Criscienzo, F. Caputo,  M. Marconi, I. Musella ]{M. Di Criscienzo$^{1}$ $^{2}$\thanks{E-mail: dicrisci@na.astro.it (MDC); caputo@mporzio.astro.it (FC); marcella@na.astro.it (MM); ilaria@na.astro.it (IM)}, F. Caputo$^{3}$\footnotemark[1], M. Marconi$^{1}$\footnotemark[1], I. Musella$^{1}$\footnotemark[1]\\
$^{1}$INAF-OAC,Via Moiarello 16, I-80131, Naples, Italy\\
$^{2}$Universit\'a  ``Tor Vergata'', via della Ricerca Scientifica, 1, Rome, Italy\\
$^{3}$INAF-OAR, Via di Frascati 33, I-00040, Monte Porzio Catone, Italy}

\begin{document}

\date{Accepted ???. Received ???; in original form ???}

\pagerange{\pageref{firstpage}--\pageref{lastpage}} \pubyear{2005}

\maketitle

\label{firstpage}

\begin{abstract}
We test whether the peak absolute magnitude $M_V(TO)$ of the Globular
Cluster Luminosity Function (GCLF) can be used for reliable
extragalactic distance determinations. Starting with the luminosity
function of the Galactic Globular Clusters listed in Harris catalog,
we determine $M_V(TO)$ either using current calibrations of the
absolute magnitude $M_V(RR)$ of RR Lyrae stars as a function of the
cluster metal content [Fe/H] and adopting selected cluster samples. We
show that the peak magnitude is slightly affected by the adopted
$M_V(RR)$-[Fe/H] relation, with the exception of that based on the
revised Baade-Wesselink method, while it depends on the criteria to
select the cluster sample. Moreover, grouping the Galactic Globular
Clusters by metallicity, we find that the metal-poor ([Fe/H] $<-$1.0,
$\langle$[Fe/H]$\rangle \sim -$1.6) sample shows peak magnitudes
systematically brighter by about 0.36 mag than those of the metal-rich
([Fe/H] $>-$1.0, ($\langle$[Fe/H]$\rangle \sim -$0.6) one, in
substantial agreement with the theoretical metallicity effect
suggested by synthetic Globular Cluster populations with constant age
and mass-function. Moving outside the Milky Way, we show that the peak
magnitude of the metal-poor clusters in M31 appears to be consistent
with that of Galactic clusters with similar metallicity, once the same
$M_V(RR)$-[Fe/H] relation is used for distance determinations. As for
the GCLFs in other external galaxies, using Surface Brightness
Fluctuations (SBF) measurements we give evidence that the luminosity
functions of the blue (metal-poor) Globular Clusters peak at the same
luminosity within $\sim$ 0.2 mag, whereas for the red (metal-rich)
samples the agreement is within $\sim$ 0.5 mag even accounting for the
theoretical metallicity correction expected for clusters with similar
ages and mass distributions.  Then, using the SBF absolute magnitudes
provided by a Cepheid distance scale calibrated on a fiducial distance
to LMC, we show that the $M_V(TO)$ value of the metal-poor clusters in
external galaxies is in excellent agreement with the value of both
Galactic and M31 ones, {\it as inferred by a RR Lyrae distance scale
referenced to the same LMC fiducial distance}. Eventually, adopting
$\mu_0$(LMC)=18.50 mag, we derive that the luminosity function of
metal-poor clusters in the Milky Way, M31, and external galaxies peak
at $M_V(TO)$=$-$7.66$\pm$0.11 mag, $-$7.65$\pm$0.19 mag and
$-$7.67$\pm$0.23 mag, respectively.  This would suggest a 
value of $-$7.66$\pm$0.09 mag (weighted mean), with any modification
of the LMC distance modulus producing a similar variation of the GCLF
peak luminosity.

\end{abstract}

\begin{keywords}
Stars, variable; clusters, globular.
\end{keywords}

\section{Introduction}

In several fields of modern astronomy, the determination of
extragalactic distances is based on a ladder which is firmly anchored
to Classical Cepheids and RR Lyrae stars, the "primary" standard
candles for Pop. I and Pop. II stellar systems, respectively, with the
properties of these variables used to calibrate "secondary" indicators
which step-by-step lead us through the Local Group up to
cosmologically significant distances.

\begin{table*}
 \centering
 \begin{minipage}{140mm}
  \caption{Globular Clusters in the Milky Way. Columns (1)-(4) are
taken from Harris (1996, 2003 update). while columns (5)-(8) give
the cluster absolute integrated magnitude according to the
$M_V(RR)$-[Fe/H] relations discussed in the text (see also Fig.2). 
Clusters marked ($a$) and ($b$) are suspected to belong to the
Sculptor and the Canis Major dwarf galaxy respectively, while Pal
1, N2419 and N5139 ($\omega$ Cen) might be associated with now not
extant dwarf galaxies (this table is available entirely in the 
electronic form).}
  \begin{tabular}{{lccccccc}}
  \hline
\hline
Name &[Fe/H] &$M_V(GC)$ &$R_{GC}$ &$M_V(GC)$
&$M_VGC)$ &$M_V(GC)$ &$M_V(GC)$\\
 (1) & (2)  &  (3:H96)    &(4:H96)    &(5:S93) &(6:F98) &(7:G03) &(8:B03)\\
\hline 
N104 & $-$0.76 &        $-$9.42 & 7.4  &        $-$9.40 & $-$9.28       & $-$9.39 &     $-$9.36 \\
N288 & $-$1.24 &        $-$6.74 & 12.0 &        $-$6.78 & $-$6.62       & $-$6.73 &     $-$6.76 \\
N362 & $-$1.16 &        $-$8.41 & 9.4  &        $-$8.45 & $-$8.29       & $-$8.41 &     $-$8.42 \\
N1261& $-$1.35 &        $-$7.81 & 18.2 &        $-$7.87 & $-$7.70       & $-$7.82 &     $-$7.85 \\
Pal 1& $-$0.60 &        $-$2.47 & 17.0 &        $-$2.42 & $-$2.32       & $-$2.42 &     $-$2.37 \\
AM 1 & $-$1.80 &        $-$4.71 & 123.2&        $-$4.84 & $-$4.62       & $-$4.75 &     $-$4.81 \\
\hline 
\end{tabular}
\end{minipage}
\end{table*}
In this context, the Globular Cluster Luminosity Function (GCLF) is
playing an ever increasing role to estimate the distance to galaxies
within $\sim$ 30 Mpc, as witnessed by the huge amount of relevant
papers published in the last decade. In the past, its use was hampered
by the lack of observations of Globular Clusters (GC) beyond the Local
Group but with the advent of modern telescopes, above all the Hubble
Space Telescope (HST), it is now possible to resolve stellar
populations in faraway galaxies, identify the GC candidates, measure
their integrated magnitude and finally build the related luminosity
function.

\begin{figure}
\psfig{figure=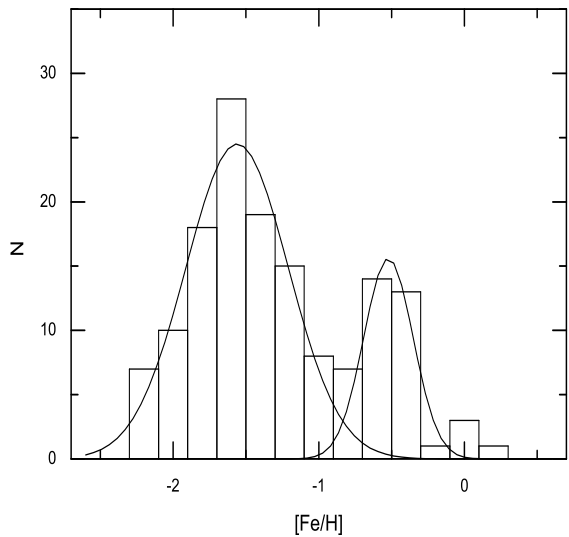,width=9cm,height=9cm}
\caption{Frequency distribution of 
metallicity for GCs in the Milky Way. The data 
have been fitted with the two Gaussian curves shown in the figure.}
\end{figure}
The GCLF method is based on the assumption that within each galaxy
hosting statistically significant numbers of GCs, the frequency of the
cluster integrated magnitude $V(GC)$ exhibits a universal shape which
can be fitted with a Gaussian distribution

$$\frac{dN}{dV}=Ae^{- \frac{[V(GC)-V(TO)]^2}{2\sigma^2}}$$
\noindent
where $dN$ is the number of clusters in the magnitude bin $dV$,
$V(TO)$ is the magnitude of the peak or turnover, $\sigma$ is the
Gaussian dispersion and $A$ the normalization factor. Once the
turnover absolute value $M_V(TO)$ is known to be constant or varying
in a predictable way, the distance to the parent galaxy follows
immediately from the apparent (reddening corrected) magnitude of the
GCLF peak. This relation is not universally accepted (see, e.g.,
Richtler 2003 and references therein) and several authors prefer to
use a $t$-distribution (see Secker 1992, Secker \& Harris 1993,
Barmby, Huchra \& Brodie 2001, hereafter BHB), but with unimportant
differences with the Gaussian turnover magnitude (Della Valle et
al. 1998, BHB). Moreover, it should be noted that also the use of the
GCLF for trustworthy distance determinations is argued because the
absolute peak magnitudes suggested so far by the various authors show
a scatter of about 0.5 mag (see Ferrarese et al. 2000). In any case,
the only galaxy where Globular Clusters can be observed well over the
turnover, down to the faintest integrated magnitudes, and where the
cluster individual distances are determined with a sufficiently high
level of confidence, as inferred from the observed magnitude of the
horizontal branch (HB) or the RR Lyrae stars, is the Milky Way. For
these reasons, the absolute LF of Galactic Globular Clusters (GGCs)
represents the first (obligatory) step to the calibration of
extragalactic luminosity functions. Unfortunately, for the Milky Way
itself current $M_V(TO)$ values show a large scatter, from $\sim -$7.3
mag (Secker 1992) to $\sim -$7.6 mag (Sandage \& Tammann 1995), thus
implying unpleasant uncertainties on the determination of the distance
to external galaxies. In order to investigate the source of such a
discrepancy, in the first part of Section 2 we estimate the effects on
the Milky Way GCLF as due to the adopted metallicity calibration of
the RR Lyrae absolute magnitude and to selective criteria of the GC
sample, while Section 3 deals with GCs in M31. As for other external
galaxies where no RR Lyrae stars are observed, in Section 4 we compare
the apparent magnitude of the GCLF turnover with the Surface
Brightness Fluctuations measurements. In this way, we also check the
consistency between GCLF distances, which are based on the RR Lyrae
luminosity scale, and those provided by the latter method, which is
calibrated on Cepheid distances. The conclusions close the paper.

\section[]{The Milky Way absolute GCLF}

\begin{table*}
\centering
\begin{minipage}{140mm}
\caption{RR Lyrae-based intrinsic distance moduli $\mu_0$(mag)
of LMC and M31. The errors in parenthesis take into account the uncertainty on
[Fe/H].}
\begin{tabular}{lccccccc}

\hline \hline
Ref. & $V_0(RR)$ &  [Fe/H] &$\mu_0$(H96) &$\mu_0$(S93)&$\mu_0$(F98)
& $\mu_0$(G03)&$\mu_0$(B03)\\
\hline
{\it LMC}\\
Wa92          &18.95(0.04)&$-$1.9&18.44(0.05)&18.58(0.07)&18.35(0.09)&18.48(0.09)&18.55(0.06)\\
Cl03          &19.06(0.06)&$-$1.5&18.49(0.08)&18.57(0.11)&18.38(0.09)&18.50(0.09)&18.55(0.11)\\
Da04          &           &$-$1.7      &           &           &           &           &18.52(0.12)\\
Bo04          &           &$-$1.5     &           &           &           &           &18.48(0.08)\\
{\it mean} & & &18.47$\pm$0.08&18.58$\pm$0.11&18.37$\pm$0.11&18.49$\pm$0.13&18.53$\pm$0.11\\
\hline
{\it M31}\\
Br04& 25.03(0.01)& $-$1.6 &24.47(0.05)&24.57(0.09)&24.37(0.06)&24.49(0.07)&24.55(0.08)\\
Br04& 25.06(0.01)& $-$1.3 &24.46(0.05)&24.51(0.09)&24.34(0.06)&24.46(0.07)&24.49(0.08)\\
{\it mean}& & &24.47$\pm$0.07&24.54$\pm$0.11&24.36$\pm$0.08&24.48$\pm$0.08&24.52$\pm$0.10\\
\hline
\end{tabular}

Ref. Wa92: Walker (1992, Globular Clusters); Cl03: Clementini et
al. (2003, Field); Da04: Dall' Ora et al. (2004, Globular Cluster, 
$K$ magnitudes); Bo04: Borissova et al. (2004, Field, $K$
magnitudes); Br04: Brown et al. (2004, Field. The two measures
refer to $ab$ and $c$-type variables. The original [Fe/H] values
are increased by 0.1 dex to put the Zinn \& West (1995) scale in
agreement with the H96 scale.)
\end{minipage}
\end{table*}

Almost all the recent papers dealing with the LF of Galactic Globular
Clusters adopt the data collected by Harris (1996). Using this catalog
(2003 update, hereafter H96), and leaving out those for which all the
required information are not available, we list in Table 1 the 144
clusters with measured metal content [Fe/H], apparent magnitude of the
horizontal branch $V(HB)$ and apparent integrated magnitude
$V(GC)$. In this Table, we have excluded AM4 whose available
photometry (Inman \& Carney 1987) shows no stars brighter than the
main-sequence turnoff. Moreover, following recent suggestions (see van
den Bergh 2003, van den Bergh \& Mackey 2004 and references therein),
we mark the clusters suspected to be not true members of the Galaxy
but of the Sculptor dwarf galaxy [N6715(M54), Ter 7, Ter 8, Arp 2, Pal
12, N4147, and Pal 2] or of the Canis Major dwarf galaxy [N1851,
N1904, N2298, and N2808]. Let us also note that the same authors
suggest that Pal 1, N5139($\omega$ Cen), and N2419 might have formed
in now disrupted dwarf galaxies.

\begin{figure}
\psfig{figure=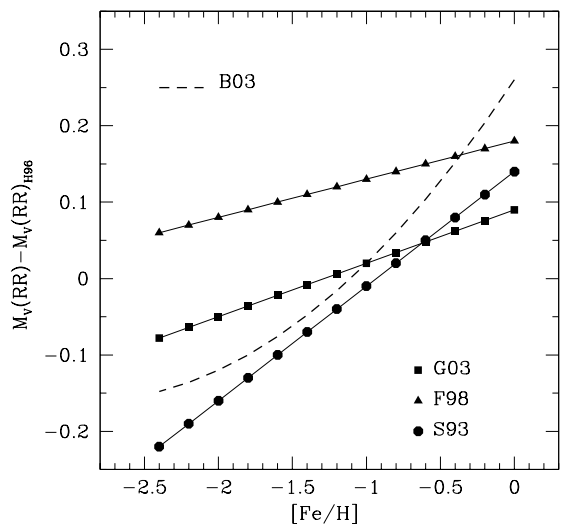,width=10cm,height=10cm}
  \caption{Comparison between the $M_V(RR)$-[Fe/H] 
relations discussed 
in the text.}
\end{figure}

The global features of the GGCs have been extensively studied (see,
e.g., van den Bergh \& Mackey 2004, van den Bergh 2003 and references
therein) and here we wish only to draw attention to the cluster
metallicity dichotomy at [Fe/H] $\sim-$1.0, with the metal-poor
component containing about 3/4 of all clusters. Based on the H96 metal
contents, we show in Fig. 1 that the total distribution can well be
described by two Gaussian curves peaked at [Fe/H] $\sim
-$1.55$\pm$0.04 ($\sigma$=0.35$\pm$0.08) and $-$0.55$\pm$0.06
($\sigma$=0.38$\pm$0.09). As a whole, no metal-rich cluster is
observed at Galactocentric distances $R_{GC} >$ 8 kpc, except the
suspected peculiar (see above) clusters Pal 1, Pal 12, and Ter 7,
while those located within 8 kpc span a metallicity range from
[Fe/H]$\sim -$2.3 to $\sim$ 0.  As for the absolute integrated
magnitude $M_V(GC)$ and $R_{GC}$ distance listed in the Harris'
catalog [columns (3) and (4) in Table 1], they rest on the cluster
distance modulus determined by adopting $V(HB)$=$V(RR)$ and the H96
relation: $$M_V(RR)=0.80+0.15[Fe/H]\eqno(2)$$
\noindent which provides a rather smooth luminosity decrease from
metal-poor to metal-rich clusters. However, the recent review by
Cacciari \& Clementini (2003) shows that a general consensus on the
$M_V(RR)$-[Fe/H] calibration of RR Lyrae stars has not been
achieved yet, with the longstanding debate concerning both the
zero point and the slope of the calibration. Since the $M_V(GC)$
values depend on the cluster distance modulus, i.e. on the adopted
$M_V(RR)$-[Fe/H] relation, we show in Fig. 2 the comparison
between equation (2) and some relevant results presented in the
recent literature. Specifically, we use the typical "long-scale"
calibration by Sandage (1993, S93)$$M_V(RR)=0.94+0.30[Fe/H],\eqno(3)$$

\noindent
the revised Baade-Wesselink ("short-scale") one by Fernley et
al. (1998, F98) $$M_V(RR)=0.98+0.20[Fe/H],\eqno(4)$$

\noindent and the relation inferred by Gratton et al. (2003, G03)

$$M_V(HB)=0.89+0.22[Fe/H]\eqno(5)$$

\noindent on the basis of the main-sequence fitting procedure.
Furthermore, since several observational and theoretical studies (see
Bono et al. 2003, Di Criscienzo, Marconi \& Caputo 2004 and references
therein) suggest that the $M_V(RR)$-[Fe/H] is not linear, becoming
steeper when moving toward larger metal content, the two linear
relations presented by Bono et al. (2003, B03) for GCs with [Fe/H]
$<-$1.6 and $\ge -$1.6 have been approximated in the quadratic form

$$M_V(RR)=1.06+0.44[Fe/H]+0.05[Fe/H]^2,\eqno(6)$$

\noindent
as shown in the figure with a dashed line. As irony of fate, all these
relations yields for the ``prototype'' variable RR Lyr itself
([Fe/H]=$-$1.39) an absolute magnitude which is consistent with the
value $M_V$=0.61$\pm$0.12 mag determined from the HST astrometric
parallax $\pi_{HST}$=3.82$\pm$0.20 mas and current uncertainty on the
extinction correction (see Benedict et al.  2002), thus hindering us
from any a priori selection.  This also in consideration of the fact
that the absolute magnitude of RR Lyrae stars is expected to depend
also on the HB morphology, becoming brighter up to $\sim$ 0.1 mag, at
fixed metal content, when the population of HB stars moves from red to
blue (see, e.g., Demarque et al. 2000, Cassisi et al. 2004 and
references therein). On this ground, we estimate that the zero-point
of all the $M_V(RR)$-[Fe/H] relations has an intrinsic uncertainty of
about 0.05 mag.

However, with everything else being constant, inspection of data in
Fig. 2 discloses that the effects of the adopted $M_V(RR)$-[Fe/H]
relation on the $M_V(GC)$ magnitude of individual clusters may amount
to quite significant values. We therefore decide to use all the
$M_V(GC)$ values listed in Table 1 to construct the LFs generated by
the various $M_V(RR)$-[Fe/H] calibrations adopted in this paper.
Before proceeding, we give in Table 2 the RR Lyrae-based intrinsic
distance moduli $\mu_0$ of the Large Magellanic Cloud (LMC) and M31,
as inferred by these $M_V(RR)$-[Fe/H] relations. We also list the
results provided by recent near-infrared observations of LMC RR Lyrae
stars and theoretical predictions discussed in B03. According to the
data in Table 2, the adopted $M_V(RR)$ calibration modifies the RR
Lyrae distance to LMC and M31, but without effect on the relative
distance of the two galaxies which turns out to be
$\mu_0$(M31)$-\mu_0$(LMC)=6.0$\pm$0.1 mag. Concerning the absolute
distance to LMC, which is a benchmark to the Cepheid distance scale,
we recall the wide range spanned by current estimates (see Caputo et
al. 2000; Gibson et al. 2000, Clementini et al., 2003), including
those provided by SN1987A ($\mu_0$=18.50$\pm$0.05 mag, Panagia 1998)
and eclipsing binaries ($\mu_0$=18.23-18.53 mag, Fitzpatrick et
al. 2003).

\begin{table}
\centering
\caption{$M_V(TO)$ and $\sigma$ values of GGCs as based on the
$M_V(RR)$-[Fe/H] relations discussed in the text. The results are
based on the H96 catalog of GGCs with (a) denoting the full
sample and (b) the Secker (1992) selection (see text).}
\begin{tabular}{lcccc}
\hline \hline
Sample & $\langle$[Fe/H]$\rangle$ & $M_V(RR)$ & $M_V(TO)$ & $\sigma$\\
\hline
H96(a) &$-$1.29$\pm$0.57 &H96  &$-$7.40$\pm$0.09& 1.11$\pm$0.12\\
N=144&                 &S93  &$-$7.46$\pm$0.11& 1.14$\pm$0.13\\
       &                 &F98  &$-$7.26$\pm$0.08& 1.11$\pm$0.10\\
       &                 &G03  &$-$7.40$\pm$0.11& 1.14$\pm$0.12\\
       &                 &B03  &$-$7.40$\pm$0.09& 1.14$\pm$0.11\\
\hline
H96(b) &$-$1.39$\pm$0.51 &H96  &$-$7.58$\pm$0.11& 1.00$\pm$0.11\\
N=100&                 &S93  &$-$7.66$\pm$0.11& 1.04$\pm$0.12\\
       &                 &F98  &$-$7.47$\pm$0.10& 1.00$\pm$0.11\\
       &                 &G03  &$-$7.62$\pm$0.11& 1.02$\pm$0.12\\
       &                 &B03  &$-$7.64$\pm$0.12& 1.00$\pm$0.11\\
\hline
\end{tabular}
\end{table}

\begin{figure}
\psfig{figure=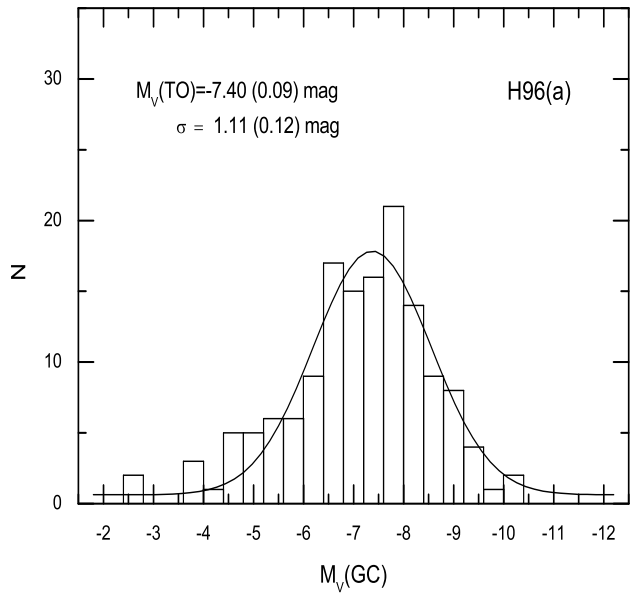,width=10cm,height=10cm}
  \caption{GCLF for Galactic clusters in our full sample H96(a).}
\end{figure}
\begin{table}
\centering
\caption{As in Table 3. but for metal-rich (MR:
[Fe/H]$>$$-$1.0) and metal-poor (MP: [Fe/H] $<$$-$1.0)
GGCs.}
\begin{tabular}{lcc}
\hline \hline 
H96(a) & MR: N=44        &MP: N=100    \\
       & $\langle$[Fe/H]$\rangle$=$-$0.57$\pm$0.26&$\langle$[Fe/H]$\rangle$=$-$1.61$\pm$0.30 \\     
\hline
$M_V(RR)$ & $M_V(TO)$($\sigma$) & $M_V(TO)$($\sigma$)\\
H96&$-$7.20$\pm$0.18(1.08$\pm$0.23)&$-$7.49$\pm$0.09(1.09$\pm$0.11)\\
S93&$-$7.17$\pm$0.19(1.08$\pm$0.23)&$-$7.56$\pm$0.10(1.12$\pm$0.12)\\
F98&$-$7.04$\pm$0.14(1.09$\pm$0.19)&$-$7.35$\pm$0.08(1.10$\pm$0.10)\\
G03&$-$7.18$\pm$0.20(1.07$\pm$0.25)&$-$7.50$\pm$0.09(1.10$\pm$0.10)\\
B03&$-$7.08$\pm$0.19(0.99$\pm$0.23)&$-$7.55$\pm$0.09(1.12$\pm$0.12)\\
\hline 
H96(b) & MR: N=26        &MP: N=74         \\
       & $\langle$[Fe/H]$\rangle$=$-$0.67$\pm$0.21&$\langle$[Fe/H]$\rangle$=$-$1.64$\pm$0.31\\
\hline
$M_V(RR)$ &  & $M_V(TO)$($\sigma$)\\
H96& $\sim-$7.4 &$-$7.63$\pm$0.09(1.00$\pm$0.10)\\
S93& $\sim-$7.3&$-$7.72$\pm$0.10(1.02$\pm$0.12)\\
F98& $\sim-$7.2&$-$7.52$\pm$0.12(1.00$\pm$0.11)\\
G03& $\sim-$7.3&$-$7.65$\pm$0.11(1.00$\pm$0.12)\\
B03& $\sim-$7.3&$-$7.70$\pm$0.11(1.00$\pm$0.11)\\
\hline
\end{tabular}
\end{table}

\begin{figure}
\psfig{figure=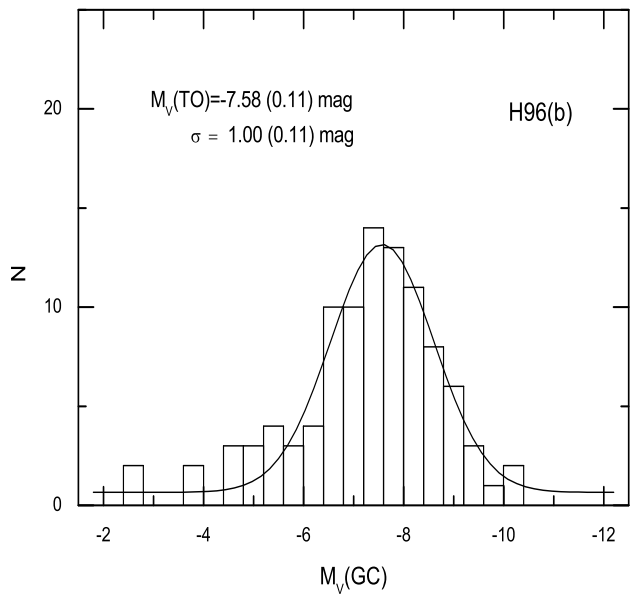,width=10cm,height=10cm}
  \caption{As in Fig. 3, but for the selected sample H96(b).}
\end{figure}

Figure 3 shows the luminosity function of our GGC full sample
[hereafter H96(a)] using the absolute integrated magnitudes listed by
H96(column (3) in Table 1). The data have been fitted with a
two-parameters (turnover and dispersion) Gaussian curve varying the
width (0.2, 0.3, 0.4 mag) and the center of the magnitude bins. The
resulting averaged values of $M_V(TO)$ and $\sigma$ are reported in
the figure. This procedure should allow us to take into account the
intrinsic dispersion of the $M_V(RR)$-[Fe/H] relation as well as the
additional effects due to the uncertainty of the apparent integrated
visual magnitude $V(GC)$ and the adopted metallicity scale. We remind
that accurate integrated photometry of Galactic GCs is difficult,
especially for those located in crowded regions toward the Galactic
Center, at large distances or with low luminosity. However, most of
the $V(GC)$ values reported in the Harris catalog, as obtained from
consistent original databases and based on concentric-aperture
photometry of the clusters, are accurate till $\sim$ 0.1 mag and only
for a small number of sparse and/or faint clusters the accuracy is
worse than 0.1 mag. On the other side, the scale from Zinn \& West
(1984) used by H96 and the one from Carretta \& Gratton (1997) adopted
by G03 show a maximum discrepancy of $\sim$ 0.2 dex at intermediate
metal deficiency ($-$1.0$\le$ [Fe/H] $\le -$1.9, see Kraft \& Ivans,
2003), thus introducing a maximum uncertainty of the order of 0.03 mag
on $M_V(RR)$.

The resulting peak magnitude$M_V(TO)$=$-$7.40$\pm$0.09 mag 
is fully consistent with$-$7.44$\pm$0.15 mag and $-$7.40$\pm$0.11 mag, 
as obtained by Kavelaars \& Hanes (1997) and Harris (2001), respectively, 
on the basis of H96 catalog and $M_V(RR)$calibration. 
We therefore repeat the procedure adopting the$ M_V(GC)$ values given 
in columns (5)-(8) of Table 1, with the numerical results labelled H96(a) 
in the first part of Table 3. Quite surprisingly, we derive that, 
in spite of the different zero points and slopes, the adopted 
dependence of $M_V(RR)$ on metallicity does not modify significantly 
the peak magnitude, with the exception of the F98 relation which gives a 
value fainter by $\sim$ 0.15 mag with respect to the average 
$M_V(TO)$=$-$7.42$\pm$0.11 mag of the other calibrations. 
In this, our turnover magnitude $M_V(TO)$=$-$7.26$\pm$0.08 mag 
based on the F98 calibration agrees with$-$7.29$\pm$0.13 mag, 
as determined by Secker (1992) using a previous GC catalog 
(Harris et al. 1991) and $M_V(RR)$=1.00+0.20[Fe/H], which is 
only 0.02 mag fainter than equation (4).

\begin{figure}
\psfig{figure=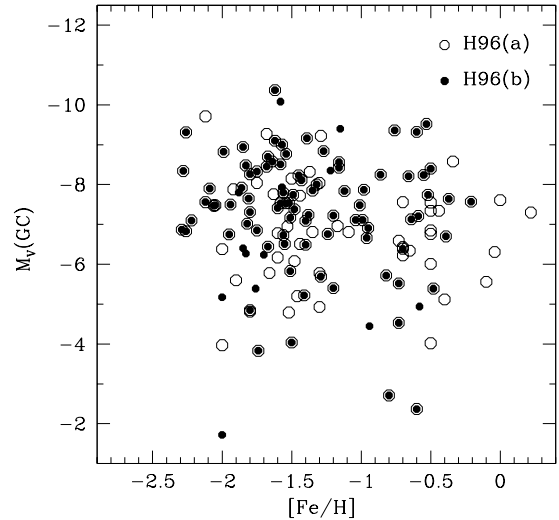,width=10cm,height=10cm}


\caption{Absolute integrated magnitudes versus Galactocentric distances for 
Galactic Globular Clusters in the full [H96(a)] and the selected [H96(b)] samples.}

\end{figure}

However, at variance with the above agreement with the quoted studies, 
our results are fainter than $M_V(TO)$=$-$7.60$\pm$0.11 mag, the peak 
magnitude obtained by Sandage \& Tammann (1995) with the S93 relation, 
and than $-$7.55$\pm$0.14 mag, as determined by Larsen et al. 
(2001, hereafter L01) from the H96 absolute integrated magnitudes. 
We note that those two results deal with the selected GGC subset 
as earlier adopted in the Secker (1992) study, namely hold for 
GCs with $E(B$-$V)\le$1.0 mag and 2 $\le R_{GC}\le$35 kpc. 
For this reason, we repeat our analysis by applying this selection 
to all the clusters in Table 1. We derive, see Fig.4 and the H96(b) 
values listed in the second part of Table 3, that the peak 
magnitudes are now brighter by $\sim$ 0.2 mag with respect to those 
of our full sample H96(a). As shown in Fig.5, where the two GC 
samples are plotted in the $M_V(GC)$-log$R_{GC}$ plane, the reason of 
such a variation is due to the fact that Secker's selection removes 
a larger number of clusters fainter than $-$7.40 mag (the peak 
magnitude of the H96(a) sample, see dashed line) with respect to 
the brighter ones, leading to the systematic increase of the peak 
luminosity. Consequently, our H96(b) magnitudes $M_V(TO)$=$-$7.66$\pm$0.11 mag 
and $-$7.58$\pm$0.11 mag, as based on equations (3) and (2), are now 
in agreement with the Sandage \& Tammann (1995) and L01 results, 
respectively, but the value based on equation (4), increased by 0.02 mag 
to account for the small difference with the relation adopted by Secker (1992), 
turns out to be significantly brighter ($\sim$ 0.16 mag) than the Secker's value.
Of importance for the following discussion is the evidence that
the constraints to select the GC sample have an effect on the
peak magnitude which may be larger than that introduced by the
adopted RR Lyrae distance scale. This is a crucial point in view
of the comparison of the Milky Way GCLF with one in another
galaxy. In particular, the fact that many external galaxies show a
bimodal metallicity distribution, as inferred from the color
behavior, and that several studies present extragalactic GCLFs
selected by the cluster metallicity or distance from the galaxy
center, lead us to analyze the dependence of the Galactic GCLF on
both [Fe/H] and $R_{GC}$.

According to Fig. 1, we split at [Fe/H]=$-$1.0 the GC full sample
and we give in the first part of Table 4 the resulting peak
magnitudes and $\sigma$ values for the metal-poor (MP) and the
metal-rich (MR) groups (see Fig. 6 which deals with absolute
integrated magnitudes based on the H96 relation). As a whole, the
peak magnitude of the MP clusters
($\langle$[Fe/H]$\rangle$=$-$1.61) is brighter by about 0.10 mag
than that of the combined sample
($\langle$[Fe/H]$\rangle$=$-$1.29) listed in the first part of 
Table 3, independently of the adopted $M_V(RR)$ calibration.
Moreover, even though the number of MR clusters is slightly
smaller than that required to measure the Gaussian parameters with
reasonable precision (N$\ge$ 50, according to BHB), 
the TO magnitude of the metal-rich
($\langle$[Fe/H]$\rangle$=$-$0.57) sample is fainter by about 0.36
mag than the value of the metal-poor one, again independently of
the adopted $M_V(RR)$ calibration. As for the Secker's selection
(i.e., H96(b) sample), we derive quite similar results, with the
peak magnitude of MP clusters ($\langle$[Fe/H]$\rangle$=$-$1.64)
brighter by about 0.34 mag and 0.05 mag than that of the few MR
ones ($\langle$[Fe/H]$\rangle$=$-$0.67) and the combined sample
($\langle$[Fe/H]$\rangle$=$-$1.39), respectively. It is worth
noticing that such an empirical evidence is consistent, also on a
quantitative way, with the theoretical calculations by Ashman,
Conti \& Zepf (1995, hereafter ACZ) who suggest a metallicity
effect $\Delta M_V(TO)$=0.32$\Delta$[Fe/H], as inferred by
synthetic cluster populations with different metallicity and
constant age and mass function.

\begin{figure}
\psfig{figure=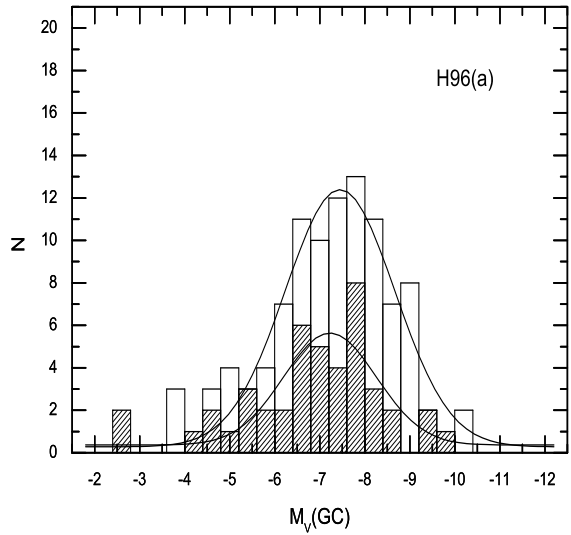,width=9cm,height=9cm}
\caption{GCLFs for metal-poor (white area) and metal-rich (dashed area) Galactic clusters in the full sample H96(a).}
\end{figure}

Furthermore, the above results are in agreement with previous
observations by Whitmore et al. (1995: M87), Kundu \& Whitmore
(1998: NGC3115), and Puzia et al. (1999: NGC 4472\footnote{For
this galaxy, Lee, Kim \& Geisler (1998) and Lee \& Kim (2000) find
little, if any, difference in the peak luminosities of blue and
red clusters.})  who find a difference between the LFs of blue
(metal-poor) and red (metal-rich) GCs in a given galaxy in the
sense that the peak visual magnitude of the former clusters is
$\sim$ 0.13, $\sim$ 0.16, and $\sim$ 0.51 mag, respectively,
brighter than that of the red ones. Moreover, L01 in their study
of relatively nearby early-type galaxies which exhibit a clear
dichotomy between blue and red GCs show that, fitting the
luminosity functions of the two populations separately, the
$V$-band turnover of the blue GCs is brighter by about 0.55 mag 
and 0.26 mag than that of the
red ones and of the combined
samples, respectively (see
the data listed in the following Table 6). In summary, also GCs in
external galaxies showing well distinct red and blue GC
populations suggest that the peak magnitude becomes fainter with
increasing the metal content of the GC sample, apparently
following the ACZ theoretical metallicity effect. We will come
back on this issue in the following section.

Concerning the dependence of $M_V(TO)$ on the cluster distance
from the Galactic center, a subdivision of metal-poor clusters
into inner halo ($R_{GC}\le$ 8 kpc) and outer halo ($R_{GC}>$8
kpc) discloses that the shape of the GCLF varies (it is broader
for the outer halo), but with no significant variation on the peak
luminosity with respect to the value of the combined sample. In
this, our result agrees with that obtained by Kavelaars \& Hanes
(1997) who use an older version of Harris catalog. For the sake
of the following discussion, we have also adopted a dividing line at
3.8 kpc for the metal-poor sample,  
but again without finding significant variation between
the peak magnitude of innermost and outermost clusters.

\begin{table}
\begin{center}
\caption{$M_V(TO)$ for metal-poor ([Fe/H]=$-$1.57) GCs in M31.}
\begin{tabular}{lcccc}
\hline \hline
$M_V(RR)$ & $\mu_0(RR)$ & $M_V(TO)$ & $M_V(TO)_{Z}$ &  $M_V(TO)_{Z}$\\
 (1)      &  (2)              & (3)   &(4) &(5)   \\
\hline
H96& 24.47 &$-$7.63$\pm$0.17 &$-$7.54 &$-$7.57   \\
S93& 24.54 &$-$7.70$\pm$0.19 &$-$7.61 &$-$7.64   \\
F98& 24.36 &$-$7.52$\pm$0.18 &$-$7.43 &$-$7.46 \\
G03& 24.48 &$-$7.64$\pm$0.18 &$-$7.55 &$-$7.58 \\
B03& 24.52 &$-$7.68$\pm$0.19 &$-$7.59 &$-$7.62   \\
\hline
\end{tabular}
\end{center}
(1): $M_V(RR)$-[Fe/H] relations discussed in Section 2; (2): RR Lyrae-based
intrinsic distance moduli from Table 2; (3) $M_V(TO)$ without
metallicity correction; (4) metallicity corrected $M_V(TO)$ for
$\Delta$[Fe/H] $\sim -$0.3 dex with respect to the average
metallicity of the H96(a) sample of GCs in the Milky Way. Errors as in column (3); 
(5) as
in column (4), but for $\Delta$[Fe/H] $\sim -$0.2 dex with respect
to the average metallicity of the H96(b) sample.
\end{table}

\section{Globular Clusters in M31}
In the field of distance determinations, the Andromeda galaxy plays a
role of great importance because it contains either Classical Cepheids
and RR Lyrae stars which provide independent distances and
consequently a valuable test for consistency between these primary
distance scales. We have shown in Table 2 that the RR Lyrae-based
distance to M31 depends on the adopted $M_V(RR)$-[Fe/H] calibration
but, for each given relation, the relative distance with respect to
LMC is constant, i.e., $\mu_0$(M31)$-\mu_0$(LMC)= 6.0$\pm$0.1 mag. It
follows that the M31 Cepheid distance $\mu_0$=24.44$\pm$0.1 mag
(Freedman \& Madore 1990) calibrated on $\mu_0$(LMC)=18.50 mag agrees
with the RR Lyrae-based value, thus providing a first evidence about
the internal consistency of the two distance scales. As for the GCLF
method, the published values of $M_V(TO)$ span a rather discomforting
range, as reported by BHB in the their recent analysis of M31 GCs.
According to these authors, who study several subsamples of the
cluster population, for the halo and disk clusters the peak magnitude
is $V_0(TO)$=16.84$\pm$0.11 mag and 16.67$\pm$0.16 mag, respectively,
while splitting the full sample at the metallicity [Fe/H]=$-$1.0 the
metal-poor ($\langle$[Fe/H]$\rangle$=$-$1.57) and metal-rich
($\langle$[Fe/H]$\rangle$=$-$0.61) groups show
$V_0(TO)$=16.84$\pm$0.16 mag and 16.43 $\pm$0.27 mag, respectively.
Moreover, a quite significant dependence of $V(TO)$ on the projected
galactocentric distance is observed: adopting a dividing line at
$R_{gc} \sim$ 3.8 kpc the innermost and outermost clusters of the
whole sample show $V_0(TO)$=16.37$\pm$0.21 mag and 16.80$\pm$0.14 mag,
while using only metal-poor clusters the peak magnitude is
16.32$\pm$0.21 mag (inner) and 17.02$\pm$0.22 mag (outer), with almost
no difference in the mean metallicity of the two subsets. As a whole,
such variations of the GCLF parameters with either $R_{gc}$ or [Fe/H]
appear at odds with the GGC behavior presented above, neither have
been reported for other galaxies. As stated by BHB, a definitive
answer on this issue will require better and less contaminated data on
the M31 clusters and for this reason we prefer to use in the following
discussion only the results concerning the full sample of metal-poor
clusters.

Using $V_0(TO)$=16.84$\pm$0.16 mag together with the distance moduli
given in Table 2, we derive the $M_V(TO)$ values listed in column (3)
of Table 5. The comparison with the Galactic values listed in the
previous Table 3 shows that the M31 absolute peak magnitudes are
brighter by about 0.25 mag and 0.04 mag than the results based on the
H96(a) and H96(b) sample, respectively. By accounting for the
metallicity correction suggested by ACZ (see values in columns (4) and
(5) of Table 5), the difference with the H96(a) results decreases to
$\sim$ 0.16 mag, while that with the H96(b) ones is almost zero.  On
the other hand, we can keep away from any metallicity effect by
directly comparing the metal-poor clusters in M31 with those in the
Milky Way.  From data in Table 4 and in column (3) of Table 5, one
derives that the M31 peak magnitudes are 0.14 mag systematically
brighter than the H96(a) ones, but almost coincident with those
inferred from the H96(b) sample. To give a reason for these results,
we note that the full sample in the BHB study is composed by clusters
out of $R_{gc} \sim$ 1 kpc from the center of M31 and that the median
galactocentric distance of the metal-poor set is 5.5 kpc, in fair
agreement with the constraints of Secker' selection which indeed was
originally thought to simulate the Galaxy as if it were viewed from
the outside and for this reason provides a better agreement with
observations of GCs in external galaxies. However, it should be
mentioned that the M31 distance moduli given in Table 2 refer to a
field population of RR Lyrae stars and that the M31 GCs are expected
at a variety of distances. In summary, no firm conclusion can be
given, although we find evidence that the LFs of Galactic and M31
Globular Clusters suggest quite similar $M_V(TO)$ magnitudes, provided
that the same $M_V(RR)$ calibration and internally consistent
constraints to select the GC samples are adopted.


\begin{table*}
\centering
\begin{minipage}{110mm}
\caption{Turnover magnitude and metallicity for GCLFs in Larsen et
al. (2001) galaxies.}
\begin{tabular}{lcccccc}
\hline \hline galaxy &  $V(TO)_{all}$ &[Fe/H]$_a$ & $V(TO)_{blue}$
&[Fe/H]$_b$
&$V(TO)_{red}$ & [Fe/H]$_r$\\
\hline
N0524   &   24.51   (0.09)& $-$0.79 & 24.34 (0.12)&   $-$1.26 &24.68 (0.14)&  $-$0.28 \\
N1023   &   23.53   (0.31)& $-$0.99 &22.82 (0.47)&   $-$1.59 &23.92 (0.39)&  $-$0.40 \\
N3115   &   22.55   (0.21)& $-$1.06 &22.45 (0.27)&   $-$1.54 &22.66 (0.33)&  $-$0.45 \\
N3379   &   22.78   (0.22)& $-$0.80 &22.57 (0.30)&   $-$1.34 &23.02 (0.32)&  $-$0.38 \\
N3384   &   23.30   (0.12)& $-$0.88 &22.98 (0.12)&   $-$1.44 &24.37 (0.30)&  $-$0.19 \\
N4365   &   24.37   (0.16)& $-$0.86 &24.01 (0.14)&   $-$1.26 &24.83 (0.23)&  $-$0.30 \\
N4406   &   23.38   (0.11)& $-$0.91 &23.28 (0.14)&   $-$1.24 &23.52 (0.17)&  $-$0.49 \\
N4472   &   23.78   (0.13)& $-$0.73 &23.38 (0.15)&   $-$1.44 &24.21 (0.23)&  $-$0.19 \\
N4473   &   23.66   (0.12)& $-$0.96 &23.46 (0.15)&   $-$1.47 &23.86 (0.15)&  $-$0.43 \\
N4486   &   23.50   (0.06)& $-$0.69 &23.36 (0.10)&   $-$1.40 &23.58 (0.07)&  $-$0.24 \\
N4494   &   23.40   (0.11)& $-$1.24 &23.24 (0.13)&   $-$1.64 &23.76 (0.22)&  $-$0.69 \\
N4552   &   23.32   (0.16)& $-$0.93 &23.01 (0.21)&   $-$1.39 &23.61 (0.24)&  $-$0.36 \\
N4594   &   22.09   (0.10)& $-$0.73 &21.80 (0.19)&   $-$1.46 &22.22 (0.12)&  $-$0.30 \\
N4649   &   23.58   (0.08)& $-$0.76 &23.46 (0.13)&   $-$1.39 &23.66 (0.11)&  $-$0.20 \\
\hline
\end{tabular}
\end{minipage}
\end{table*}

\begin{table*}
\centering
\begin{minipage}{110mm}
\caption{SBF magnitudes and differences with the GCLF turnover magnitudes 
for Larsen et
al. (2001) galaxies.}
\begin{tabular}{lcccc}
\hline \hline 
galaxy &     $m_i^*$  & $m_I^*-V(TO)_{all}$ & $m_I^*-V(TO)_{blue}$ & $m_I^*-V(TO)_{red}$\\				
\hline
N0524	&     30.16	(0.20) & 5.65 (0.22) &	5.82 (0.23)	   &	5.48 (0.24)	\\				
N1023	& 	28.55	(0.15) & 5.02 (0.34) &	5.73 (0.49)	   &	4.63 (0.42)	\\				
N3115	& 	28.19	(0.08) & 5.64 (0.22) &	5.74 (0.28)	   &	5.53 (0.34)	\\				
N3379	& 	28.38	(0.10) & 5.60 (0.24) &	5.81 (0.32)	   &	5.36 (0.33)	\\				
N3384	& 	28.59	(0.13) & 5.29 (0.18) &	5.61 (0.18)	   &	4.22 (0.33)	\\				
N4365	& 	29.82	(0.16) & 5.45 (0.23) &	5.81 (0.21)	   &	4.99 (0.28)	\\				
N4406	& 	29.43	(0.13) & 6.05 (0.17) &	6.15 (0.19)	   &	5.91 (0.21)	\\				
N4472	& 	29.31	(0.09) & 5.53 (0.16) &	5.93 (0.17)	   &	5.10 (0.25)	\\				
N4473	& 	29.24	(0.12) & 5.58 (0.17) &	5.78 (0.19)	   &	5.38 (0.19)	\\				
N4486	& 	29.30	(0.15) & 5.80 (0.16) &	5.94 (0.18)	   &	5.72 (0.17)	\\				
N4494	& 	29.43	(0.09) & 6.03 (0.14) &	6.19 (0.16)	   &	5.67 (0.24)	\\				
N4552	& 	29.19	(0.13) & 5.87 (0.21) &	6.18 (0.25)	   &	5.58 (0.27)	\\				
N4594	& 	28.21	(0.17) & 6.12 (0.20) &	6.41 (0.26)	   &	5.99 (0.21)	\\
N4649	& 	29.39	(0.14) & 5.81 (0.16) &	5.93 (0.19)	   &	5.73 (0.18)	\\
\hline																			
{\it mean}	&	       & 5.67 (0.31) &	5.93 (0.21)    &	5.38 (0.50)	\\
\hline
\end{tabular}
\end{minipage}
\end{table*}

\section{External galaxies}In their paper, ACZ show that the theoretical metallicity correction on the peak magnitude helps to remove the discrepancy
between the GCLF and the Surface
Brightness Fluctuations (SBF) distance scales. 
Following a different approach, we note that in the case of
galaxies for which both types of methods are possible the GCLF
universality can straightway be tested by considering the
difference between $V(TO)$ and $m^*$, the SBF magnitude adjusted
to a fiducial color (see later). Since the $m^*$ absolute
calibration is assumed to depend only on a zero-point, such a
difference provides information for or against the
constancy of the GCLF peak absolute magnitude, independently of
the galaxy distance. With such a purpose, in the following we
adopt the $V(TO)$ magnitudes measured by L01 
by two-parameters fits to GCLFs in early-type galaxies
together with the correspondent $I$-band SBF
measurements by Tonry et al. (2001,
hereafter T01) as adjusted to the fiducial color $(V-I)_0$=1.15
mag according to the T01 relation

$$m_I^*=m_I-4.5[(V-I)_0-1.15]\eqno(7)$$

\noindent
where $(V-I)_0$ is the galaxy color.

\begin{table*}
\centering
\begin{minipage}{100mm}
\caption{SBF calibration from Cepheid distances.}
\begin{tabular}{@{}lccccc@{}}
\hline \hline
galaxy & [O/H] & $m_I^*$ & $\mu_0$(KP$_n$)& $\mu_0$(KP$_{n,Z}$)& $\mu_0$(F02) \\
\hline
LMC   & $-$0.40  &             & 18.50         &   18.50            &     18.50      \\
N7331   & $-$0.23& 28.86 (0.14)& 30.81 (0.09)  &   30.84 (0.09)     &   30.71 (0.09)    \\
N3031   & $-$0.15& 26.21 (0.25)& 27.75 (0.08)  &   27.80 (0.08)     &   27.63 (0.09)    \\
N4258   & $-$0.05& 27.57 (0.08)& 29.44 (0.07)  &   29.51 (0.07)     &   29.32 (0.06)    \\
N4725   &   +0.02& 28.87 (0.32)& 30.38 (0.06)  &   30.46 (0.06)     &   30.28 (0.07)    \\
N0224   &   +0.08& 22.67 (0.05)& 24.38 (0.05)  &   24.48 (0.05)     &   24.30 (0.08)    \\
N3368   &   +0.30& 28.34 (0.20)& 29.97 (0.06)  &   30.11 (0.06)     &   30.09 (0.10)    \\
N4548   &   +0.44& 29.68 (0.53)& 30.88 (0.05)  &   31.05 (0.05)     &   31.20 (0.05)    \\
\hline
galaxy  &        &             & $M_I^*$       &     $M_I^*$        &  $M_I$   \\
N7331   &        &           & $-$1.96 (0.17)  &  $-$1.99 (0.17)  & $-$1.86 (0.17) \\
N3031   &        &           & $-$1.54 (0.26)  &  $-$1.59 (0.26)  & $-$1.42 (0.26)  \\
N4258   &        &           & $-$1.87 (0.11)  &  $-$1.94 (0.11)  & $-$1.75 (0.10) \\
N4725   &        &           & $-$1.51 (0.33)  &  $-$1.59 (0.33)  & $-$1.40 (0.33)  \\
N0224   &        &           & $-$1.71 (0.07)  &  $-$1.81 (0.07)  & $-$1.64 (0.09)  \\
N3368   &        &           & $-$1.63 (0.21)  &  $-$1.77 (0.21)  & $-$1.75 (0.22)  \\
N4548   &        &           & $-$1.20 (0.53)  &  $-$1.37 (0.53)  & $-$1.52 (0.53) \\
\hline
{\it median}  & & & $-$1.63 (0.05) &  $-$1.77 (0.05)  & $-$1.64 (0.05)  \\{\it w-mean}  & & & $-$1.75 (0.05) & $-$1.84 (0.05) & $-$1.68 (0.05)\\\hline 
\end{tabular}
\end{minipage}
\end{table*}

All the galaxies studied by L01 exhibit clear bimodal color
distribution and for such a reason the peak magnitude was measured for
either the blue and the red populations, as well as for the combined
samples. These magnitudes are reported in Table 6 together with the
mean metallicity of the combined, blue, and red samples, as determined
using the L01 intrinsic $(V-I)_0$ colors and the relation of Kundu \&
Whitmore (1998), while Table 7 gives the T01 $m_I^*$ values.  As shown
in the upper panel of Fig. 7, which deals with the combined samples
and where open circles refer to measured peak magnitudes while filled
ones depict the metallicity corrected values scaled to the average
value [Fe/H]=$-$1.3 of all the GGCs (see Table 3), the difference
between the two magnitudes is $m_I^*-V(TO)$=5.67$\pm$0.31 mag (no
metallicity correction, dashed line) and 5.81$\pm$0.31 mag
(metallicity corrected, solid line), thus suggesting for the GCLFs in
these galaxies a reasonably similar absolute peak magnitude.

Moreover, by considering the blue clusters separately, we show
in the lower panel in the same figure that the peak magnitudes
scale even better with the SBF measurements yielding a difference
$m_I^*-V(TO)$=5.93$\pm$0.21 mag (no metallicity correction, dashed line) and
5.99$\pm$0.22 mag (solid line) with a correction that
accounts for the difference between the [Fe/H] values of the blue
clusters (see data in column (5) of Table 6)
and the average metallicity of the
Galactic metal-poor clusters ([Fe/H]=$-$1.6, see Table 4). As for the red
clusters, we find that the two magnitudes are poorly correlated
for we derive $m_I^*-V(TO)$=5.38$\pm$0.50 mag (no metallicity
correction) and 5.46$\pm$0.48 mag (metallicity corrected to the
average metal content [Fe/H]=$-$0.6 of metal-rich GGCs (see Table 4)).

\begin{figure}
\psfig{figure=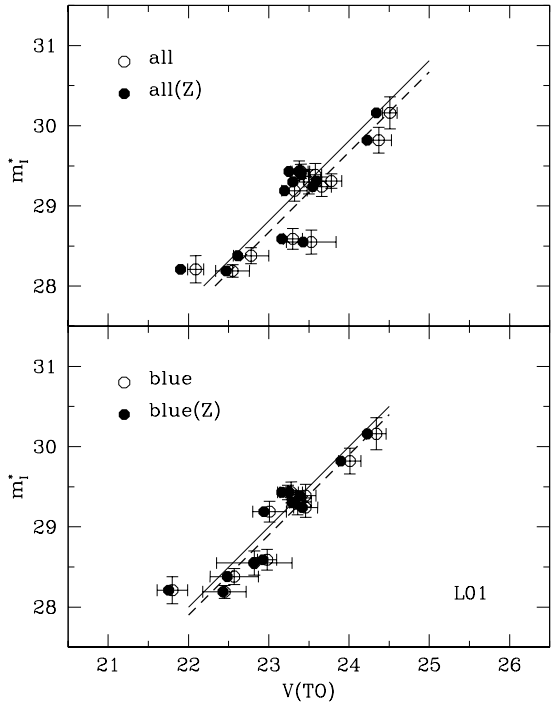,width=10cm,height=10cm}
  \caption{SBF measurements 
versus GCLF peak magnitudes for the external galaxies studied 
by Larsen et al. (2001). The upper panel refers to the combined 
samples of GCs, while the lower one deals with blue (metal-poor) 
clusters. Open and filled circles depict observed and metallicity 
corrected peak magnitudes, respectively (see text).}
\end{figure}

In summary, provided that the $m_I^*$ absolute calibration is
assumed to rest on a zero-point, the L01 data
for galaxies showing a bimodal
distribution in the GC colors suggest that the luminosity 
functions of the blue (metal-poor) clusters  
peak at the same absolute magnitude within $\sim$ 0.2 mag, while 
for the GCLFs of the combined samples  
the constancy is attained within $\sim$ 0.3 mag
as a result of the quite scattered peak magnitudes of
the red (metal-rich)
globular clusters. This behavior holds even if the ACZ 
theoretical metallicity
correction is taken into account, likely suggesting that 
in external galaxies
the metal-rich GCs may have different ages and/or mass distributions
than
the metal-poor component.
Apparently, this result disagrees with Kundu \& Whitmore (2001a,b)
whose sample contains
few galaxies which  show evidence of bi-modality in the GC color
distribution, but for  which they find agreement between the GCLF
and the SBF distances considering the metal content and peak
magnitude of the GC full samples. However, by inspection
of Table 6 of Kundu \& Whitmore (2001a) we note that the difference
$\Delta\mu_0$(GCLF-SBF\footnote{SBF data from Neilsen 1999})
varies from $-$0.27 to +0.67 mag, while from Table 3 in
Kundu \& Whitmore (2001b) one has that the difference between the
GCLF distance moduli and those from the literature ranges from
$-$0.70 to +0.38 mag, depending on the galaxy.

We stress again that all the above discussion relies on the assumption
that the absolute calibration of the SBF $m_I^*$ magnitudes depends
only on a zero-point. In this case, the differences
$m_I^*-V(TO)_{blue}$ given in Table 7 would yield that the absolute
peak magnitudes for the metal-poor clusters in the L01 galaxies have a
scatter of about 0.2-0.3 mag, which means 2-3 Mpc at Virgo distances.
As a matter of the fact, looking at the $M_I^*$ magnitudes given by
Tonry et al. (1999) for six calibrating galaxies (see their Table 2)
one finds a range of $\sim$ 0.5 mag, with even the best SBF
measurements giving $M_I^*=-1.77\pm 0.12$ (NGC224) and $-2.04\pm 0.19$
mag (NGC7331). On this ground, it is quite difficult to distinguish
whether the scatter in the $m_I^*-V(TO)_{blue}$ differences reflects a
real scatter of the absolute peak magnitudes or is due to the
intrinsic uncertainty of the SBF calibration.
 
Concerning the latter point, Tonry et al. (1999), using the Ferrarese
et al. (2000) HST Cepheid distances to the six calibrating galaxies,
prefer to adopt the median value $M_I^*=-$1.74$\pm$0.08 mag rather
than the weighted mean $-$1.80$\pm$0.08 mag, given the wide range in
the errors in the SBF measurement. Accordingly, for all the blue
clusters in the L01 galaxies we derive $M_V(TO)$=$-$7.67$\pm$0.23 mag
and $-$7.73$\pm$0.23 mag, depending on whether the ACZ metallicity
correction is neglected or used, respectively, {\it within a Cepheid
distance scale calibrated on $\mu_0$(LMC)=18.50 mag.}

This is a crucial point to be considered before comparing these 
peak absolute magnitudes with those of metal-poor clusters 
in the Milky Way since the values listed in Table 4 are 
correlated to the LMC distance moduli given in Table 2 for the 
various $M_V(RR)$-[Fe/H] relations. In 
other words, if we adopt $\mu_0$(LMC)=18.50 mag, 
then the peak absolute magnitude of
Galactic metal-poor clusters is $M_V(TO)$=$-$7.50$\pm$0.10
mag and $-$7.66$\pm$0.11 mag for the H96(a) and H96(b) samples,
respectively. On this basis, as already presented for the
metal-poor GCs in M31,
we find a better agreement with the Galactic peak magnitude
dealing with the Secker's selection of GCs.

However, as discussed in Jensen et al. (2003), the zero-point of
the SBF calibration follows the uncertainties of the Cepheid
scale: indeed, we show in Table 8 that the revised HST Cepheid distances
determined by Freedman et al. (2001, KP$_n$) for the SBF calibrating
galaxies\footnote {For N224 (M31) ground observations by Freedman \& Madore
(1990) were used.}
would lead to the slightly fainter median value
$M_I^*=-$1.63$\pm$0.05 mag and weighted mean $-$1.75$\pm$0.05 mag. 
Furthermore, one should also consider the occurrence of a
metallicity effect on the Cepheid distance scale. Using the
empirical relation adopted by Freedman et al. (2001), namely
$\Delta \mu_0=-$0.2$\Delta$[O/H] where $\Delta$[O/H] is the
difference between the oxygen abundance of the galaxy and that of
LMC, the metallicity corrected distance moduli (KP$_{n,Z}$) yield
a median value and a weighted mean as 
$M_I^*=-$1.77$\pm$0.05 mag and $-$1.84$\pm$0.05 mag, respectively.

\begin{figure}
\psfig{figure=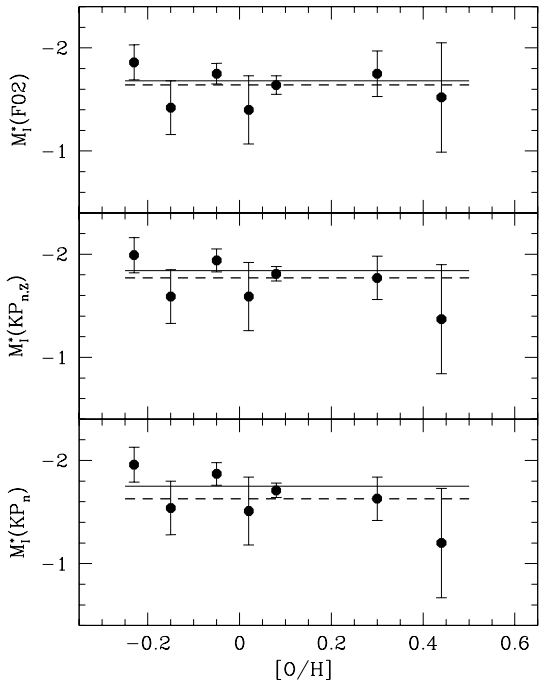,width=10cm,height=10cm}
 \caption{SBF absolute magnitudes of calibrating galaxies as a function 
of the galaxy oxygen abundance. The three panels deal with the 
revised Cepheid distances by Freedman et al. (2001) 
without metallicity correction (KP$_n$) and adopting either 
the empirical correction (KP$_{n,Z}$) or the theoretical one (F02). 
Dashed and solid lines show the median value and the weighted mean, respectively.}
\end{figure}

In several papers, the occurrence of a metallicity effect on the
Cepheid distance scale is rejected also in consideration of a
mistakenly believed disagreement between the empirical correction
adopted by Freedman et al. (2001) and the predicted one $\Delta
\mu_0$=+0.27$\Delta$log$Z$, as based on nonlinear convective 
models of Cepheid structures (see Caputo, Marconi \& Musella 2002
and references therein) with $Z$ in the range of 0.004 (Small
Magellanic Cloud) to 0.02 (roughly solar chemical composition). As
a matter of fact, it should firstly be clear that the empirical
correction holds with the oxygen abundance of the parent galaxy,
whereas the theoretical one is based on the chemical composition
of the Cepheids. Moreover, it has been shown (Fiorentino et al.
2002, F02) that the theoretical correction is not linear over the whole
metallicity range covered by galaxies hosting Cepheids, with a
turnover at about solar chemical composition and with sign and
amount of the correction depending on both the helium and metal
content of the Cepheid. On this basis, F02 showed that 
the empirical metallicity
correction suggested by Cepheid observations in two fields of the
galaxy M101 may be accounted for adopting a helium-to-metal
enrichment ratio $\Delta Y/\Delta Z \sim$ 3.5, as also confirmed
on the basis of an updated extended model set (Marconi, Musella 
\& Fiorentino 2005). 
It is also of interest to note that recent
high-resolution spectroscopic abundances for Galactic and
Magellanic Cloud Cepheids (Romaniello et al. 2005) show that the
Cepheid luminosities are incompatible with the empirical linear
correction, whereas are fairly described by the F02 non-monotonic
theoretical behavior with a
helium-to-metal enrichment ratio $\Delta Y/\Delta Z$=2.5-3.5.
As for the effects on the SBF calibration, we show in Fig. 8 the
absolute $M_I^*$ values of the calibrating galaxies as a function
of the [O/H] abundance of the galaxies (see also Table 8). The
three panels refer to the HST revised distance moduli by Freedman et al.
(2001) without metallicity correction (KP$_n$) and using both the
empirical (KP$_{n,Z}$) and the theoretical correction (F02) with
$\Delta Y/\Delta Z$=3.5. With reference to the median values 
(dashed line) and the weighted 
means (solid line),   
one has that a
metallicity correction to the measured Cepheid distances is {\it needed} to
remove the trend of $M_I^*$ with the oxygen abundance, and that 
the theoretical relation seems to work better than the empirical
one to give a fairly constant SBF zero-point.  

Eventually, we note that using the weighted mean  $M_I^*$=$-$1.68$\pm$0.05 mag provided
by the theoretically corrected distance moduli to
the SBF calibrating galaxies together with the ACZ metallicity 
correction yields that the peak absolute magnitude of 
the L01 blue GCs is $M_V(TO)=-$7.67$\pm$0.23
mag which is astonishingly coincident with the values
$-$7.66$\pm$0.11 mag and $-$7.65$\pm$0.19 mag inferred by  metal-poor clusters
in the Milky Way and M31, respectively, {\it at $\mu_0$(LMC)=18.50 mag.}
In closure, we wish to mention that a recent theoretical SBF calibration 
(Cantiello et al. 2003) yields $M_I^*$=$-$1.74$\pm$0.23 mag, as determined 
using stellar evolutionary tracks computed with the same input physics 
adopted for our pulsation 
models of RR Lyrae and Cepheid structures, in particular for those computed 
by B03 and F02. On these grounds, we should adopt $\mu_0$(LMC)=18.53 mag and 
the weighted mean provided by the SBF calibrating galaxies becomes  
$M_I^*$=$-$1.71$\pm$0.05 mag, that is practically coincident with the theoretical 
value, with  
the above $M_V(TO)$ luminosities increased by 0.03 mag. 
     
\section{Conclusions}

In this paper, we have investigated the universality of the GCLF and the
use of the peak magnitude for reliable
distance determinations to external galaxies. 
The main results may be summarized as follows:
\begin{enumerate}
\item   Concerning the dependence
of the Milky Way GCLF on the adopted $M_V$-[Fe/H] relation to get the
cluster distances, we find no significant effects on the 
absolute peak magnitude $M_V(TO)$, with 
the exception of the one
based on the revised
Baade-Wesselink method (F98), that provides a fainter magnitude 
by about 0.15 mag. Moreover, we show that the selection of
the GC sample may influence the peak magnitude: in particular, for 
each given $M_V(RR)$-[Fe/H] relation, using 
only GCs with reddenings $E(B-V)\le$ 1.0 mag and Galactocentric distances 
2$\le R_{GC}\le$ 35 kpc, as earlier suggested by Secker (1992) to treat 
the Galaxy as if it were viewed from the outside, yields that 
the peak magnitude  
becomes systematically 
brighter by about 0.2 mag.  
As a whole, the combined effects of the adopted $M_V(RR)$ calibration and   
selective criteria are the main reason for the discordant Galactic peak magnitudes 
presented in the relevant literature.        

\item Grouping the Galactic clusters by metallicity, 
the peak magnitude of the
metal-poor ([Fe/H]$ <-$1.0] subsample is brighter than 
that of the metal-rich ([Fe/H]$>-$1.0] one by about
0.36~mag. This empirical results meets, also in a quantitative way, 
the theoretical metallicity effects suggested by
Ashman, Conti \& Zepf (1995) on the basis of synthetic GC populations 
with similar age and mass-function. 
As for the dependence on the Galactocentric
distance, we found that the shape of the GCLF is broader for the outer
halo ($R_{GC}>8$ kpc) than for the inner one, but with no significant
effect on the peak luminosity.

\item Using BHB data for metal-poor GCs in M31, 
we find a close agreement with 
the metal-poor Galactic sample results,  as obtained 
according to the Secker's
selection and using 
the same $M_V(RR)$ calibration to get cluster distances. 

\item Concerning 
external galaxies with available deep photometry
and close enough to have apparent GCLF extending 
below the turnover, we use the sample provided by Larsen et al. (2001) 
which contains 
galaxies showing a bimodal distribution of the GC color
(and consequently of the metallicity). Given the absence of 
RR Lyrae stars to measure the galaxy distances, we use the $I$-band SBF
measurements (Tonry et al. 2001) to evaluate the
difference between the apparent peak magnitude $V(TO)$  
and the SBF magnitude $m_I^*$, as adjusted to the fiducial
color $(V-I)_0=1.15$~mag. In this way, we show that the blue 
(metal-poor) cluster
component peaks at the same luminosity 
within $\sim 0.2$~mag, while the GCLFs of the full  
samples show constant values 
within $\sim 0.3$~mag as a consequence of the 
quite scattered peak magnitudes of the red globular clusters. 
The adoption of the theoretical
metallicity correction by ACZ does not significantly
modify these results, thus suggesting that in external galaxies 
blue and red globular clusters may have different ages and/or mass 
distributions.

\item Following the universally accepted assumption 
that the absolute calibration of the SBF $m_I^*$ magnitude depends only on a
zero-point, we analyze the Cepheid distances to the calibrating 
galaxies, as determined by Freedman et al. (2001) within a Cepheid  
distance scale calibrated on
$\mu_0(LMC)=18.50$~mag. We firstly show that the SBF absolute 
magnitude $M_I^*$ of the calibrating galaxies becomes brighter 
with decreasing the galaxy oxygen abundance, suggesting the 
occurrence of a metallicity effect on the Cepheid distance scale. 
Once the Cepheid distances are corrected using either the empirical 
(Freedman et al. 2001)  
or the theoretical (Fiorentino et al. 2002) metallicity corrections, 
the trend is reduced. In particular,  we find that the peak absolute magnitude of the 
extragalactic metal-poor clusters 
is practically identical to 
the Milky Way and M31 values, 
provided that the Secker's selection of Galactic clusters and the theoretical 
metallicity corrections on both the GCLF peak magnitude and the 
Cepheid distance are adopted. 
\item 
Finally, {\it within a Cepheid and RR Lyrae distance scale calibrated
on $\mu_0$(LMC)=18.50 mag}, the three sets of metal-poor GCs give
$M_V(TO)$=$-$7.66$\pm$0.11 mag (Milky Way), $-$7.65$\pm$0.19 mag
(M31), and $-$7.67$\pm$0.23 mag (extragalactic clusters). This would
suggest a  value of $-$7.66$\pm$0.09 mag (weighted mean),
with any modification of the LMC distance modulus producing a similar
variation of the GCLF peak luminosity.

\end{enumerate}

\section*{Acknowledgments}
Financial support for this study was provided by MIUR under the
PRIN project ``Continuity and Discontinuity in the Milky Way
Formation'' (P.I. Raffaele Gratton) and by Regione Campania under the
scientific project ``Fundamental Astronomy: refining the steps of the
distance ladder'' (P.I. Marcella Marconi).
It is a pleasure to thank G. Bono, V. Castellani, E. Iodice,
S. Larsen and  B. Madore for several comments, and E. Brocato
and G. Raimondo for discussions on the SBF method.



\end{document}